\documentclass{article} 
\usepackage{anysize}
\newenvironment{mylist}{\begin{list}{--}{\topsep=6pt \parskip=0pt \itemsep=0pt \leftmargin=0.5cm}}{\end{list}}
\newenvironment{otherlist}{\begin{list}{$\bullet$}{\topsep=6pt \parskip=0pt \itemsep=0pt \leftmargin=0.5cm}}{\end{list}}
\newenvironment{mynum}{\begin{enumerate}{\topsep=0pt \parskip=0pt \itemsep=0pt \leftmargin=0.5cm}}{\end{enumerate}}

\title{Women in Astronomy Workshop}

\author{Sarah Brough$^1$\thanks{E-mail: sb@aao.gov.au (workshop chair)}, Amanda E. Bauer$^1$, Kate Brooks$^2$, Andrew Hopkins$^1$, Sarah Maddison$^3$\\
$^1$ Australian Astronomical Observatory, PO Box 296, Epping, NSW 1710, Australia\\
$^2$CSIRO Astronomy \& Space Science, Australia Telescope National Facility (ATNF),\\
PO Box 76, Epping NSW 1710, Australia\\
$^3$Centre for Astrophysics \& Supercomputing, Swinburne University, H39, PO Box 218,\\ 
Hawthorn, VIC 3122, Australia\\
}

\begin{document}

\maketitle

\begin{abstract}
\noindent Here we report on the Women in Astronomy Workshop\footnote{\texttt{http://asawomeninastronomy.org/meetings/wia2011/}}, which was held on 13 May 2011 in Sydney, Australia. The workshop was organised by the Astronomical Society of Australia's Chapter on Women in Astronomy, to discuss some of the issues that face women in astronomy and make recommendations to help support the success of women in Australian astronomy but we came to broader conclusions that have value for the whole astronomical community. The workshop consisted of four sessions, with presentations by invited speakers on demographics, leadership, varied career paths, and how institutions \& individuals can help.The workshop ended with a discussion panel that summarised the day's debate and presented a list of recommendations for the Australian astronomical community (both individuals and institutions) that are provided in this report.  
\end{abstract}

\vspace{-0.5cm}
\section{Introduction}

Astronomy, as with other Physical Sciences, is a field in which gender is not balanced, with fewer women represented at each increasing level of seniority.  A major goal of this workshop was to raise awareness of issues that face women which may impede their career progress or cause them  to drop out of the field, as well as propose solutions that can be applied by institutions and individuals to retain women in the profession and help them succeed.      The workshop focused on the following four goals:
\begin{mynum}
\item To raise awareness of issues that impact on the decision of women to continue in a research focused career.
\item To illustrate the validity of the many career paths for astronomers and space scientists.
\item To identify issues faced by women in astronomy and some solutions.
\item To suggest how institutions and individuals can help women succeed in our field.
\end{mynum}

The workshop was, of course, open to both women and men. Many of the issues that affect the career progress of women also affect men, and so removing barriers that women face will improve the environment for all astronomers.  Furthermore, given that the majority of senior astronomers are men who through their positions of influence can implement changes, the participation of men in the workshop was vital.

We invited a number of astronomers to speak at the workshop and address these goals, as well as a professional speaker on women's issues.  The speakers presented their experiences from both personal and managerial viewpoints.  At the end of the day we discussed the issues that had been raised and presented some practical solutions that institutions and individuals can implement.

\section{Demographics}

The good news is that more than half the students attaining undergraduate science degrees in Australia are women. Of those that choose to pursue a PhD, roughly 50\% are women. Within a few years of getting a PhD, however, the percentage of participating women starts dropping dramatically.  It is incredibly poor business practice to invest in the training of so many women only to lose all their knowledge, expertise, and training within a few short years.  

{\bf Sarah Maddison} (Swinburne University) presented the statistics currently available for Australia.  Snapshot surveys clearly indicate relatively high levels of participation at undergraduate and post-graduate levels within Australia, but persistently low levels of representation of women at senior levels. The 2007 DEEWR\footnote{DEEWR is the Australian Department of Education, Employment and Workplace Relations} Selected Higher Education Student Statistics shows that women make up 55\% of all undergraduate students in Australia and about 52\% of postgraduate students.  Within the Natural and Physical Sciences in Australia,  women comprise 47\% of research doctorate completions, just under  30\% of the senior lecturers (level C) within academia, and about 12\% of academics above the senior lecturer (Bell, 2009).  
Focussing on Australian astronomy, the National Committee of Astronomy (NCA)\footnote{The National Committee of Astronomy (NCA) is one of the committees of the Australian Academy of Science, {\tt http://www.aao.gov.au/nca/}} decadal and mid-term reviews show that the number of women PhD students has increased from 15\% in 1996 to 40\% in 2005, but has stayed flat in the past five to 2010, while the total number of PhD students has increased by almost 50\% in that same 5 year period.  The number of women academics in Australian astronomy increased from about 15\% in 1996 to 19.7\% in 2005, and has remained relatively flat in the intervening 5 years to 2010 (21.5\%), while the overall number of astronomy positions rose by 28\% in that same time (mainly due to increased government funding for astronomy and an associated increase in the number of post doctoral positions).   

The main aim of the Astronomical Society of Australia's Women in Astronomy Chapter\footnote{\tt{http://asawomeninastronomy.org/}} is to monitor the status of women working in Australian astronomy and recommend future actions for improvement. To do so, it is important that we have accurate statistical information on which to base an evaluation. 
The Chapter is working towards implementing sustainable processes for obtaining a broad range of demographic information in an ongoing fashion in cooperation with the NCA as part of the decadal and mid-term review process, as well as collecting gender statistics on the Astronomical Society of Australia's membership, its prizes, and participation in its Annual Scientific Meetings.  
As well gender and progression statistics within academia, it is important for us to collect statistics on the visibility and recognition of women through metrics such as awards and prizes, grant success, membership of high-level executive committees, and invitation to speak at conferences. These data on the recognition and esteem of women will allow us to celebrate successes, as well as highlight potential unconscious biases and implement strategies to overcome them.

\section{Leadership}

{\bf Ros Dubs} (Chair, Space Industry Innovation Council) spoke from the perspective of someone who has completed a PhD and two postdoctoral positions, but stepped away from research into a management role.  She is now Chair of the Space Industry Innovation Council as well as a member of the Australian Astronomical Observatory's new Advisory Committee.  Ros highlighted the importance of scientific training as a crucial success factor for all her management roles.  She advised that trained scientists should not ignore broader career opportunities and that there are other avenues for doing science, particularly ones where you can work towards advancing science.  She mentioned some of the paths available from higher levels of science education:  From a science degree one can take on further study, roles in public policy, program delivery, science communication, research \& development or quality management;  from an engineering degree one can take on further study, project management, R\&D, instrument development, or facility operations; a PhD opens up a research career, scientific assistant to senior management, technology transfer, strategic planning, or company operations. Finally, a PhD in addition to operational experience, opens up positions as senior adviser to government, managing major business lines and non-executive director roles.
In terms of her own career progression and choices, Ros was the first speaker to highlight the importance of mentors. 
 
Too many successful managers still believe that hard work and merit are the only factors necessary for advancement.  They have ignored the systematic barriers that disadvantage certain groups (e.g. women) for too long.  Now that the positive impact on profits of having a diverse workforce, and avoiding retraining and
hiring costs, is becoming clearer, there is incentive to change.  It is important to survey reasons for women dropping out, as well as highlight initiatives that have worked. Universities and businesses are beginning to offer conference travel money, childcare assistance, extended allowances and acknowledgement of career breaks for publications record, teaching relief and research assistant funding for women returning from maternity breaks.

Our professional speaker {\bf Nicole McKenna} (Principal, McKenna Bespoke Development \& Facilitation, Vice President; National Foundation for Australian Women) highlighted the importance of borrowing learnings and
interventions from sectors outside astronomy to inform the work we do on gender equality in the profession. She brought a fresh view to the table and highlighted the importance of recognising the role played by unconscious bias in affecting women's career progress in astronomy and beyond. Nicole began with a retrospective look at the achievements of the women's movement over time, and related that to the current challenges faced by women wanting to pursue careers.  She then presented some of the strategies leaders and managers can use to support and develop their staff and set them up for success.

In her discussion of unconscious bias, Nicole explored the idea that we all have preferences in terms of the sorts of people with whom we like to work and socialize.  The challenge is in recognising these preferences, and then taking steps to put them to one side in circumstances where there is value in working with someone who doesn't fall into our preferred zone.  She suggested that leaders and upstream managers must make conscious choices more often to support, develop and offer opportunities to women in their team if we are to stem the tide of women opting out.  While gender equity of itself is a worthy basis for such an approach, she reinforced that diverse teams typically deliver better outcomes and that women opting out of intellectual work for which they are well educated and trained is a huge waste of an organisation's investment in them over many years. Some interesting statistics were shared, including the impact of blind auditions on the number of women chosen for roles in orchestras. There was also some discussion by participants of the statistics around the gender make up of invited speakers at conferences versus the number of contributed speakers, which is much more gender balanced.

Nicole talked about the impact of linear career paths on women, particularly those who wish to combine family responsibilities with their careers.  She also discussed the importance of encouraging mentoring within organisations, and for women to be proactive in identifying people within their organisation who can sponsor their development and act as mentors. If there is no mentoring program in place in your institution, Nicole suggested being proactive about developing one by engaging like-minded people with the influence to make it happen.  If it is not possible to create a program within an institution, then it may be possible to use external mentors, who can often bring a freshness and independence to the table that is helpful.

Creating opportunities for women to mix with one another and with clients or stakeholders important to the organisation is also helpful.  The key is to make sure that the activities and events organised are appropriate. Organisations will sometimes resort to sports focused events or paint-balling, which is often unappealing to women.  A good mix of events that caters for broader tastes is likely to be much more successful and effective.

Nicole introduced her notion of ``pink tasks"; a phenomenon she has come across in her work over many years. These are tasks that need to be done on time and to a high standard, but where there is little substantive development or increased visibility for the person undertaking or assigned the tasks. While these tasks are necessary, she has seen too many examples of where they are assigned to women over and over again, so that they become experts in circumstances that support the business but do not raise their profile or position them better for promotion or advancement.  Nicole gave a
number of examples of the sorts of tasks that fall into this category, and warned participants to steer clear of becoming experts in these sorts of things.  Rather, she suggested that when there is an opportunity to choose between working behind the scenes doing low level research, reviewing data or taking on an extra teaching load and attending an important but maybe difficult meeting, accepting a speaking engagement that takes you out of your comfort zone (even if it makes you nervous) or otherwise collaborating with more senior people in a way that will enhance your development, make sure you choose the latter at least half of the time.  To the extent that leaders or managers recognise these sorts of things are happening, they must be courageous enough to take steps to stop them.

Some of Nicole's advice to women to help enhance their career and improve their chances of achieving their goals included the following:

\begin{mylist}

\item Recognise who in your department or institution is influential, successful, powerful and hopefully shares some common values or interests with you and get to know them! Seek their advice, counsel and mentorship.

\item Do not let people assume that just because you do a task once that you will always be responsible for it, especially if it is not offering you any development or other career advantages that will help you achieve your
goals.

\item Remind yourself that it's OK to say ``No" from time to time.  It is more important to be respected than to be liked, and being liked because you allow others to steam roll you isn't a great place to be anyway.

\item Develop some career goals that are specific, measurable, attractive to you, realistic, and time-framed (short- and long-term).

\item Women can tend not to speak up in meetings unless they are almost certain that what they are about to say is absolutely correct. Be brave, and get comfortable with the prospect that you might not be 100\% right.  Be known as a vocal contributor of substance and not a silent (but excellent) note taker.

\item There are lots of awards, prizes and promotions out there, and they can all contribute to you achieving your career goals.  The problem is that you will not get them if you do not apply for them.  Remember to back yourself.
\end{mylist}

\section{Varied career paths}

{\bf Kate Brooks} (CSIRO Astronomy \& Space Science), {\bf  Ilana Feain} (CSIRO Astronomy \& Space Science), {\bf Tanya Hill} (Melbourne Planetarium), {\bf Elaine Sadler} (University of Sydney) and {\bf Maria Cunningham} (University of NSW) shared their personal stories with us.  This is not something we are used to doing in academia and it  takes courage to share your own life experiences with your professional colleagues so we thank them wholeheartedly.  Their honesty and openness in discussing some of the challenges that they face balancing childcare with a research career were eye-opening.  It really emphasised the issues many parents (primarily women) face on a day-to-day basis.  This led to one of the many positive outcomes of this meeting.  It was apparent from Nicole and Elaine that career breaks have been hidden in the past, sometimes for fear of being discriminated against, and issues faced by women have not been raised for fear of being labelled complainers.  The simple case of creating a space where women could share stories and issues was extremely valuable.  Acknowledging that here are some issues that are not going to go away, for instance women needing to take career breaks for family, but do need to be considered and discussed openly, without prejudice, is a very important step to take towards improving the system for future generations.

While these talks all came from differing perspectives, there were common themes. First was the importance of mentors.  Each of these women felt that they would not be where they are today without the guidance, support and belief of mentors.  Secondly the importance of vacation scholarships in getting them into astronomy research was clear. They also emphasised the positives of working in research, including: 
\begin{mylist}
\item flexible work arrangements,
\item family-friendly practices,
\item paid maternity leave,
\item the collaborative nature of astronomy, and
\item skype! 
\end{mylist}
They also discussed some of the challenges: 
\begin{mylist}
\item the inability to work after hours with young families,
\item inflexible (and expensive!) daycare options,
\item limitations with working part-time, and
\item needing to keep travel to a minimum.
\end{mylist}

Each speaker highlighted the value of their partner in helping with childcare, while Elaine highlighted an issue for astronomers who work far from their family support network: the lack of hands-on grandparents or other nearby family members to help out with childcare. Maria also reminded us that some of the issues faced by women are a result of personality traits rather than gender, with some traits being more prevalent in women, for example lack of self-confidence and unwillingness to put themselves forward for awards and promotions.

\section{How individuals and institutions can help}

{\bf Lisa Harvey-Smith} (CSIRO Astronomy \& Space Science) summarised the result of the recent Women in Science and Engineering (WiSE\footnote{http://www.scienceinpublic.com.au/category/wise}) summit held in Canberra in April 2011.  The  Summit brought together science and industry leaders, advocates, and high-achieving young women in science and engineering in a high profile event to encourage science and engineering leaders to take practical steps to secure a higher return on their investment in young women scientists and engineers.  We, as a nation, are not successfully supporting their transition into independent researchers and science leaders.   The loss of these highly trained, smart women means we are not tapping into our full national innovation potential.  As a result of this, the Australian Research Council (ARC) and National Health and Medical Research Council (NHMRC) agreed to change how they assess research publications in the grant applications of those with interrupted careers.  The ARC committed to extending the period taken into account. The NHMRC will from this year consider any nominated five years of an applicantÕs career rather than simply the previous five years.   
Science \& Technology Australia, formerly FASTS\footnote{The Federation of Australian Scientific \& Technological Societies.}, will conduct an audit of practices amongst its member societies with the aim of increasing the participation of women through best practice.  FASTS will gather examples of existing practices, programs and policies which have been successful and develop a toolkit to guide the science and technology sector.

{\bf Naomi McClure-Griffiths} (CSIRO Astronomy \& Space Science) spoke about the return to active research following maternity leave with the assistance of a return-to-work grant.  The difficulties of returning to work following childbirth include reduced working hours, resulting in a drop-off in productivity; coping with the unpredictability of young children and the travel necessary in astronomy.  No one wants to leave their young child behind when traveling and sometimes it is simply not possible to do so.  Traveling with young children is not easy: it is expensive, babysitters are required, and it generally means less networking (no chats in the pub after the meeting).  The Ruby Payne-Scott award, available at the CSIRO, is a research grant of up to \$35,000 available to parents returning to work after at least 18 weeks of leave as primary carer for a newborn child.  The award is very flexible and may be used to pay for a research assistant, to buy-out from projects in order to commit time to research,  funding visitorsÕ travel, funding meetings or funding partner/nanny travel so that the recipient can attend international meetings.  Naomi used her award to pay for a 4-week visit to Europe, including her partner's travel, a 4-week trip to the USA which included funds to bring US collaborators to the National Radio Astronomy Observatory in Green Bank, West Virginia, where she was located, as well as bringing distinguished visitors to Sydney and running the ``Diffuse Gas in and Around the Milky Way" workshop, which benefited the entire Australian astronomical community. This sort of ``value-adding'' of the award was also seen recently with the recent ALMA Community Workshop that was partially funded by the Ruby Payne-Scott awards of Kate Brooks and Ilana Feain. 
Naomi mentioned some points that could be improved in the Ruby Payne-Scott award, including support staff to help organise meetings funded by the award.  While the application is simple and easy, it may be necessary to apply while still on maternity leave which is not ideal. It would be helpful to be able to spread the funding over 2 years.  Also helpful would be increased flexibility in spending the award, including being able to use the award to pay for a nanny at home and in being able to apply for the grant any time in the first 3 years back at work, as it can take time to formulate your research plan after a maternity break.

Other things that could help all parents returning to work include on-site childcare, flexible work hours (including encouragement to work from home), travel support for up to (and beyond) preschool-age children, family-friendly office space (even if babies are not at work with their mother, breast-feeding still needs to occur regularly).  The ASA's Women in Astronomy Chapter maintains a list of best practise in family-friendly policies on their website\footnote{\tt{http://asawomeninastronomy.org/action/familyfriendly/}} to help provide ideas for their members to take to their own institutes to implement.

{\bf Andrew Hopkins} (Australian Astronomical Observatory) spoke about his personal perspective on managing staff, starting with the appointment process. It is obviously important to have gender balance on the selection panel, and offering family-friendly working arrangements (flexible hours, part time options) can make your job attractive to a wider range of people.  If the applicants are not gender balanced, then managers should try to analyse why.  Perhaps the advertisement was not inclusive, possibly specific women applicants were not targeted as well as male applicants. It is important to take career breaks into account when assessing performance.  Finally, being aware of unconscious biases and ask yourself continually whether your preference for a candidate is because they are the best, or because they are the most like yourself.  

Once you have employed someone, it is your responsibility to mentor them.  You should be aware of their personal goals, career level and personal constraints.  This can be achieved through regular interaction and discussion, where encouragement and identifying ways forward can also occur.  Provide opportunities where possible for them to advance their career and empathise with their position:  Review, listen, be creative.  Andrew also raised the issue of the ``impostor syndrome," which affects most people to some degree, but more often women, and potentially to a career-debilitating degree (Kaplan 2010; Ivie \& Ephraim 2011). The impostor syndrome describes the fear or worry that eventually someone will figure out that you are not actually as smart and capable as they think you are.  This can prevent you from negotiating contracts, asking for promotions, or applying for grants or positions that you think you will not get. But you will never get something you don't apply or ask for.

{\bf Bryan Gaensler} (University of Sydney and CAASTRO Director) discussed the ways in which the new ARC Centre of Excellence for All-sky Astrophysics (CAASTRO) is aiming to be a family-friendly and gender equitable research centre from the outset.  Starting with their job advertisements, they have made it policy to offer all positions available either full-time or part-time  (although the part-time option is only available to Australian citizens and permanent residents due to visa requirements).  Of 12 research positions awarded so far, two have been taken up part-time.  When deciding on salary levels for each new appointment, they also think carefully about gender equity and pay gaps.  Every time a position is offered, it can be made at a certain level. Since men are more likely to negotiate for a higher salary, managers can unwittingly create a gender pay difference.  During new hires at CAASTRO, they are considering what they have or have not offered previously, and what they are willing or reluctant to do in the future.
CAASTRO also ensures that core meetings take place between 10am--2pm so that people picking up children from school/childcare can attend important meetings. CAASTRO will also be involved in school programs targeting girls in science and is organising a formal mentoring program for early-career researchers.  CAASTRO will encourage the nomination of PhD students and early-career researchers for prizes and awards to help bypass and overcome the impostor syndrome.

Bryan also discussed some personal strategies for managers to implement.  Stay informed about the issues that women are facing through mailing lists, policies, studies and papers, workshops (e.g. AASWOMEN\footnote{\tt{http://www.aas.org/cswa/AASWOMEN.html}}, WiSeNet\footnote{\tt{http://www.wisenet-australia.org/}}, sciencewomen\footnote{\tt{http://scienceblogs.com/sciencewoman/}} and scienceprofessor\footnote{\tt{http://science-professor.blogspot.com/}} blogs). Carefully check the adjectives you use in reference letters before sending them. Madera et al. (2009) found that women are often described as ``more communal and less agentic than menÊ[and this has] a negative relationship with hiring decisions in academiaÓ.  Communal adjectives include: kind, warm, tactful, nurturing, agreeable, caring, helpful. Agentic adjectives include: assertive, confident, ambitious, independent, outspoken, daring.  Selection panels should down-weight any reference letters that provide extraneous information. Stewart (2011) lists the following inappropriate examples: ``it's amazing how much she's accomplished"; ``it appears her health is stable"; ``she is close to my wife".  Other behaviours that Bryan highlighted that everyone should be aware of is cutting people off when they are speaking (this more often happens to women than men) and using names in inconsistent way (people often use surnames for men but first names for women).  Managers should welcome kids in the workplace if feasible, especially in school holidays, and read ``What Can Men Do To Help Women Succeed in Astronomy?''\footnote{\tt{http://www.aas.org/cswa/Jan11/townhall.html}}, which suggests a range of behaviours to aim for and to avoid!  Finally, scrutinise the gender balance at all conferences, committees and events.  

\section{Discussion \& Recommendations}

The discussion panel session at the end of the workshop aimed to provide a clear list of ideas that institutions can implement to help women remain in astronomy and succeed to more senior levels.  The discussion points included:
\begin{mylist}
\item{How much of the problem do we think is caused by unconscious bias and how can we combat that?}
\item{To what extent is mentoring a primary solution and what other alternatives are there?}
\item{What solutions can be identified in order to maximise research productivity following maternity breaks via return to work schemes?}
\end{mylist}
The recommendations for both institutions and individuals that came out of the discussion are outlined below.

\subsection{Recommendations for institutions}

\begin{otherlist}
\item{Appoint diverse committee members to select speakers for conferences, recipients of awards and jobs, and to sit on important decision-making boards.} 
\item{Offer and advertise flexible working arrangements in new positions.}
\item{Advertise your new positions widely and try to specifically target women (e.g. via the AASWomen mailing list).}
\item{Encourage flexible work arrangements in your institution.}
\item{Ensure that all important meetings are held between 10am and 2pm, and finish meetings on time!}
\item{Supply childcare at professional meetings, especially national meetings.}
\item{Offer flexible grants to encourage parents to return to research after parental leave.}
\item{Offer contract extensions equivalent to the parental leave taken.}
\item{Pay attention to adjectives in letters of recommendation, both those you write and those you receive.}
\item{Encourage employees at all levels to participate in organised mentoring programs.  If your organisation is small, there may be suitable external mentors from collaborative organisations, and there are companies who can provide mentoring.}
\item{In job/grant applications, ask for best N years of publications, rather than previous N years, to account for career breaks.}
\item{Discuss these issues in your working groups and encourage the acknowledgement of unconscious bias.}
\end{otherlist}

\subsection{Recommendations for individuals}

\begin{otherlist}
\item{Ask!  If you want something, you will not get it without asking.  You never know what will happen if you ask.}
\item{Find suitable mentors.   Seek out senior members who can be your cheer squad (both men and women).}
\item{Speak up. Make a vocal contribution of substance in order to be noticed, heard, acknowledged and appreciated.}
\item{Set goals. You drive your own career.}
\item{Know when to say no. The point is not to be liked, but to be respected.}
\item{Be self aware of personal biases.}
\item{Pay attention to adjectives when writing recommendation letters.} 
\item{When judging grant, award or job applications, take career breaks into account.}
\item{Be a mentor.  Your experience and support can be valuable to others.}
 \end{otherlist}
 
\section{Conclusion}

Fewer women are represented at each increasing level of seniority in astronomy.  The loss of these highly trained women is a waste to our field and results in a loss of diversity at the most senior levels.  The Astronomical Society of Australia's Chapter on Women in Astronomy held a workshop to discuss some of the issues that face women in astronomy and make recommendations to help support the success of women in Australian astronomy. 

The workshop was very successful.  Seventy-five people from 16 institutes across Australia registered, and the whole process was very positive.  A number of recommendations (outlined above) for both institutions and individuals were presented to help women succeed in astronomy.
We were very pleased at the reactions from senior managers who appreciated the concrete ideas that came out of the meeting, and felt that they could go away and implement many of them. On the scale of organisational budgets, most strategies that can enhance the status of women in astronomy are relatively inexpensive to implement. 

We do have one final note of caution.  Everyone who attended the workshop realises that women in astronomy face a range of barriers and are interested in working towards dismantling them.  Of those who did not attend there are people who do not think that there is a problem, who think that they are not biased so they do not have anything to gain by attending such a workshop, who think the problem is impossible to solve so why bother, and men who do not think they can get something out of such an event.  The recommendations presented at the workshop suggest otherwise, that there is a problem, that everyone is biased and should be aware of it, that there are solutions and that the recommendations benefit everyone, not just women.  With this in mind, we all need to continue discussing these issues in our institutions and at meetings to continue raising awareness.

\subsection*{Acknowledgements}
We thank the sponsors of this meeting, the Australian Astronomical Observatory (AAO), ARC Centre of Excellence for All-sky Astrophysics (CAASTRO) and CSIRO Astronomy \& Space Science (CASS).

\subsection*{References}

Bell, S. (2009). ``Women in science: maximising productivity, diversity and innovation''. Canberra: Federation of Australian Scientific \& Technological Societies (FASTS)

Ivie, R. \& Ephraim, A (2011). ``Women and the Imposter Syndrome in Astronomy'', Status, January 2011, p4.

Kaplan, K. (2010). ``Self-doubt plagues female astronomers'', Nature 463, 574 

Madera, J. M., Hebl, M. R., Martin, R. C. (2009). ``Gender and Letters of Recommendation for Academia: Agentic and Communal Differences'', J. Applied Psychology, 94, 1591

Stewart, A. (2011). ``Addressing Unconscious Bias: Steps toward an Inclusive Scientific Culture'', AAS, 217, 414.01

``Australian Astronomy: Beyond 2000. Decadal Plan 1996-2005''. Canberra: National Committee for Astronomy of the Australian Academy of Science (August 1995)

``New Horizons: A Decadal Plan for Australian Astronomy 2006-2015". Canberra: National Committee for Astronomy of the Australian Academy of Science (November 2005) 

\end{document}